\documentclass{ws1}

\usepackage{dsfont}
\usepackage{mathrsfs}

\newcommand\dd{\mathrm{d}}
\newcommand\ee{\mathrm{e}}
\newcommand{\Scri}{\mathscr{I}}
\newcommand\diff[2]{\frac{\partial#1}{\partial#2}}

\begin{document}

\markboth{J\"org Hennig and Marcus Ansorg}
{A Fully Pseudospectral Scheme for Solving Singular Hyperbolic Equations}

%
%

\title{A FULLY PSEUDOSPECTRAL SCHEME FOR SOLVING SINGULAR HYPERBOLIC
EQUATIONS ON CONFORMALLY COMPACTIFIED SPACE-TIMES}

\author{J\"ORG HENNIG\footnote{Electronic address: pjh@aei.mpg.de}
    \ and MARCUS ANSORG\footnote{Electronic address: mans@aei.mpg.de}
   }

\address{Max Planck Institute for Gravitational Physics\\
Am M\"uhlenberg 1, D-14476 Golm, Germany}

%

\maketitle


\begin{abstract}
{\bfseries Abstract.}\quad
With the example of the spherically symmetric scalar wave
equation on Minkowski space-time we demonstrate that a fully
pseudospectral scheme (i.e. spectral with respect to
both spatial {\em and} time directions) 
can be applied for solving hyperbolic equations. The calculations
are carried out within the framework of conformally compactified
space-times. 
In our formulation, the equation becomes singular at null infinity and 
yields regular boundary conditions there. In this manner it becomes possible
to avoid \lq\lq artificial\rq\rq\ conditions at
some numerical outer boundary at a finite 
distance. We obtain highly accurate
numerical solutions possessing exponential spectral convergence, a feature known from solving
elliptic PDEs with spectral methods. Our investigations are meant as a
first step towards the goal of treating time evolution problems in
General Relativity with spectral methods in space and time. 
\end{abstract}

\keywords{Wave Equation; Spectral Methods; Conformal Compactification.\\
\emph{Preprint number:} AEI-2008-033}
\section{Introduction}

The simulation of dynamical processes within the theory of
General Relativity plays an important role for understanding
astrophysical phenomena, for studying the stability of equilibrium
configurations
by introducing small perturbations and evolving in time
and for predicting the
properties of the emitted gravitational waves. An ideal way of 
analyzing such processes carefully
would be the construction of explicit solutions to Einstein's field equations.
However, due to the mathematical complexity of these equations, even
stationary configurations can be described in terms of analytic expressions
in only a few exceptional cases. Therefore, the
only chance for tackling time dependent
processes is the application of numerical methods.  
Nevertheless, in order to come as close as possible to an explicit solution, 
it is desirable to find mathematical descriptions and
numerical procedures that permit
the computation of a very accurate approximation to the solution in
question. 

As we demonstrate in this paper, a promising approach
towards this goal is the combination
of highly accurate (pseudo-) spectral methods
with the fruitful concept of conformally compactified
space-times\footnote{In principle, it is also possible to apply the
pseudospectral algorithm described in this paper to arbitrary compact
domains if
the boundary conditions at $\Scri$ are replaced by other conditions at
the boundaries of these domains.}.
By means of spectral methods, general relativistic equilibrium configurations
have been obtained with almost machine accuracy
\cite{Ansorg}.
Dynamical relativistic
problems have also been studied utilizing spectral methods
(with respect to the spatial directions, combined with finite difference
methods in time direction),
see e.g. \cite{Boyle}.
(For a comprehensive overview of the applications of spectral methods in
general relativity see \cite{Novak}.)
However, there is only little experience
regarding spectral expansions with respect to space \emph{and}
time\footnote{Only in the context of \emph{finite} and \emph{spectral
element methods}
the simultaneous space-time treatment of hyperbolic equations
is already more common, see
e.g. \cite{Uengoer} and references therein.} \cite{Ierley}. As a first step
towards the goal of treating time evolution problems in 
General Relativity with a \emph{fully} pseudospectral scheme, we study
model equations,  
in particular linear and non-linear wave equations.

The concept of \emph{conformal infinity}, i.e. conformally compactified
space-times,  
was introduced by Penrose \cite{Penrose} in 1964 
(For an overview on this topic see~\cite{Frauendiener}; numerical
studies can be found in 
\cite{Frauendiener2, Huebner, Husa, Zenginoglu}).
Within this scheme we are able to carry out the numerical calculations
up to infinity. As a consequence, we 
have no need of
numerical outer boundaries at a finite physical distance. At such finite
boundaries one usually imposes  
particular conditions in order to complete the mathematical problem.
These conditions have to be compatible with the differential equation to
be solved and should lead to a well-posed problem.
(Pseudospectral methods are particularly sensitive with
respect to this issue.  
In the worst case, the numerical simulation breaks down after some time when
the errors arising due to  incompatible boundary conditions accumulate.)
However, in any case the physical meaning of such boundary data and their
influence to the solution is unclear. That is why we prefer the
conformal compactification.

An additional important feature of the conformal approach is the precise
determination of gravitational wave signals at null infinity $\Scri$,
which is located at \emph{finite coordinate distance}. Hence one avoids
approximative wave 
extraction techniques at \emph{finite physical distance}.

Within the conformal concept, hyperbolic differential equations 
become singular at null infinity. Although it is possible to carry out
an appropriate regularization in some special cases
(for Einstein's field equations, a
suitable reformulation in terms of Friedrich's regular conformal field
equations~\cite{Friedrich} can be used), we demonstrate here that this
degeneration of the equations permits a careful numerical treatment. In
combination with a fully pseudospectral scheme, we obtain highly
accurate numerical solutions (up to 12 or 13 correct digits for a double
precision 
code). 

The paper is organized as follows. In Sec.~\ref{sec_wave_eq}, we recall
the conformal compactification of Minkowski space-time. 
Moreover we consider the scalar wave equation on this background and
derive the boundary 
conditions to be imposed at the singular points of this equation. The
numerical method 
for solving the singular equation is explained in detail in
Sec.~\ref{sec_num}. In the subsequent section, we give a number of
numerical examples in order to test the method and to study the accuracy
of the numerical solutions. In Sec.~\ref{sec_other}, we show that
the application of the
spectral scheme is not restricted to the homogeneous linear wave
equation. To this end we study two additional example equations: an
inhomogeneous wave equation and a non-linear wave equation.
In Sec. \ref{sec_regularization}, we consider a regularized version of
the wave equation and demonstrate that our numerical method is also
applicable to this equation. Finally, in Sec.~\ref{sec_discussion} we
discuss our results.  

\section{Wave Equation and Compactification}\label{sec_wave_eq}
\subsection{The wave equation}
The model equation to be studied throughout most of this
paper is the spherically symmetric wave equation 
\begin{equation}\label{wave_eq}
  \Box f \equiv f_{,rr}+\frac{2}{r}f_{,r}-f_{,tt} = 0,
  \quad f=f(r,t)
\end{equation}
on Minkowski space-time with the line element
\begin{equation}
  \dd s^2 = \dd r^2+
  r^2(\dd\vartheta^2+\sin^2\!\vartheta\,\dd\varphi^2)-\dd t^2
\end{equation}
using spherical coordinates $(r,\vartheta,\varphi, t)$.
The general solution to \eqref{wave_eq},
\begin{equation}
  f(r,t)=\frac{a(r+t)+b(r-t)}{r},
\end{equation}
with $a,b\in C^2(\mathds{R})$, can be used
to investigate the accuracy of the numerical solution. Here we restrict
ourselves to solutions which are regular at $r=0$.
As a consequence, we obtain the condition
$b(x)=-a(-x)~\textrm{for all}~x\in\mathds{R}$,
i.e. we consider solutions of the form
\begin{equation}\label{sol1}
  f(r,t)=\frac{a(t+r)-a(t-r)}{r}.
\end{equation}

\begin{figure}[ht]
  \centerline{\psfig{file=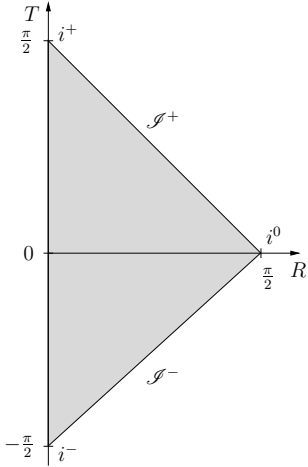,width=4cm}} 
  \vspace*{8pt}
  \caption{The standard conformal diagram of Minkowski space-time. The
  line $R=0$  corresponds to the center of symmetry $r=0$. Furthermore,
  past and future null infinity $\Scri^\pm$, past and future time-like
  infinity  $i^\pm$, and space-like infinity $i^0$ are marked accordingly.
  \label{fig_Minkowski}} 
\end{figure}

For the conformal compactification of Minkowski space-time, one often
uses the following 
coordinate transformation (see \cite{Frauendiener}):
\begin{equation}\label{coord}
  r=\frac{1}{2}\left[\tan(R+T)+\tan(R-T)\right],\quad
  t=\frac{1}{2}\left[\tan(R+T)-\tan(R-T)\right],
\end{equation} 
through which new coordinates $R$ and $T$ are introduced, with
$R\in[0,\frac{\pi}{2}]$, $ T\in[R-\frac{\pi}{2},\frac{\pi}{2}-R]$.
The line element takes the form
\begin{equation}\label{line}
  \dd s^2 = \frac{1}{Q^2 P^2}\left[4(\dd R^2-\dd T^2)
  +\sin^2(2R)(\dd\vartheta^2+\sin^2\!\vartheta\,\dd\varphi^2)\right]
\end{equation}
 with
\begin{equation}\label{QP}
 \quad Q:=\sqrt 2\cos(R+T),\quad P:=\sqrt 2\cos(R-T).
\end{equation}
Obviously, the metric components are singular at null
infinity $\Scri$, i.e. at $R\pm T
= \pi/2$ where $P=0$ or 
$Q=0$.
However, the rescaled line element $\dd\bar s^2=Q^2P^2\dd s^2$ is
regular at $\Scri$. The domain on which $R$ and $T$ are
defined is illustrated  
by the well-known standard conformal diagram, see Fig. \ref{fig_Minkowski}.
 
For the wave equation in terms of the coordinates $R$ and $T$ one obtains
\begin{equation}\label{wave_eq1}
  QP(f_{,RR}-f_{,TT})
  +2\frac{(Q^2+P^2)f_{,R}+(Q^2-P^2)f_{,T}}{\sin(2R)} = 0.
\end{equation}
The general solution \eqref{sol1} can be written as
\begin{equation}\label{sol2}
  f(R,T)=\frac{A(T+R)-A(T-R)}{\tan(R+T)+\tan(R-T)}
\end{equation}
with $A(x) := a(\tan x)$ for all $x\in [-\pi/2,\pi/2]$.
Due to the appearance of the functions $P$,
$Q$ and $\sin(2R)$, the wave equation \eqref{wave_eq1}
is singular at infinity (as a consequence of the compactification) and at the center of
symmetry (as an effect of the spherical coordinates), i.e. at the
following points: 
\begin{itemize}
  \item $i^-$ and $\Scri^-$: $R-T=\frac{\pi}{2}$\quad$\Rightarrow$\quad $P=0$
  \item $i^+$ and $\Scri^+$: $R+T=\frac{\pi}{2}$\quad$\Rightarrow$\quad $Q=0$
  \item $i^0$:  $R\pm T=\frac{\pi}{2}$\quad$\Rightarrow$\quad $P=Q=0$
  \item $R=0:$ $\sin(2R) = 0$.
\end{itemize}
In the numerical scheme we need to treat these points particularly
carefully, in order to derive well-defined 
boundary conditions, as will be discussed in the next subsection.

\begin{figure}[ht]
  \centerline{\psfig{file=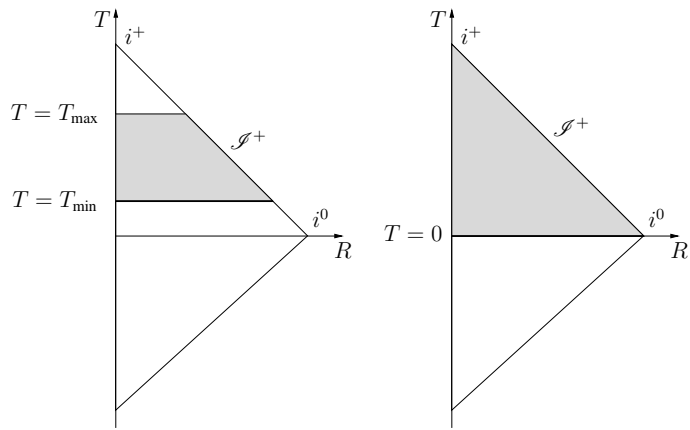,width=9cm}} 
  \vspace*{8pt}
  \caption{The numerical domains for the
  \emph{hyperboloidal initial boundary value problem} (left) and
  the \emph{standard Cauchy problem} (right).
  \label{fig_IBVP}} 
\end{figure}

We study two different types
of initial boundary value problems (IBVP)
for  the wave equation \eqref{wave_eq1}, see Fig.~\ref{fig_IBVP}. In a
\emph{hyperboloidal IBVP} we prescribe initial data
(the values of $f$ and the
time derivative $f_{,T}$) on the hyperboloidal slice\footnote{A slice is
called \emph{hyperboloidal} if it is space-like everywhere and extends
up to $\Scri$.}
$T=T_\textrm{min}$
and evolve these data up to another hyperboloidal slice
$T=T_\textrm{max}$.
For the second type we consider a \emph{\lq\lq standard\rq\rq\ Cauchy problem},
in which initial data are given on the particular Cauchy surface $T=t=0$. From these data, 
the entire future of $f$ up to $i^+$ and $\Scri^+$ is determined. As a consequence of the time symmetry
of the wave equation, the past of $f$ can be calculated in the same manner. 
We thus obtain the function $f$ \emph{everywhere}. 

For relativistic time evolution problems, initial data on \emph{Cauchy}
surfaces  have to be constructed very carefully in order to avoid
(logarithmic) singularities at $i^0$
which cannot be removed easily by means of a coordinate transformation.
Solutions to  this problem have been discussed by Corvino \cite{Corvino},
utilizing a
\lq\lq gluing\rq\rq\ technique in order to obtain initial data which are
\emph{exact Schwarzschild} data near spatial infinity, and more generally
by Dain and Friedrich \cite{Dain}.
Note that, on the other hand,
there are specific conditions which guarantee the regularity at $\Scri$
of initial data on \emph{hyperboloidal} slices \cite{Andersson1,
Andersson2}. 
For this reason, we
concentrate primarily on the hyperboloidal IBVP problem here.

\subsection{Boundary conditions}

The appropriate boundary conditions\footnote{Note that we denote any
equations to be imposed at inner and outer borders of the numerical domain
as \lq\lq boundary conditions\rq\rq,
even if they may be of quite different nature
(regularity conditions at
coordinate singularities, conditions at $\Scri$, second order
conditions, \dots).}
at the singular boundaries $\Scri$
and $R=0$  follow from an analysis of the wave equation \eqref{wave_eq1}
at these \lq\lq critical\rq\rq\ points. In this investigation we assume
regularity of the solution, 
i.e. bounded first and second order derivatives.

In order to derive a condition at $R=0$,
we first multiply \eqref{wave_eq1} with $\sin(2R)$ and then perform the
limit $R\to 0$. 
Since $P=Q=\cos T$ as $R\to 0$, the following Neumann condition 
\begin{equation}\label{BCCenter}
  f_{,R} = 0
\end{equation}
arises. Being a necessary requirement in the context of spherical symmetry,
this condition also appears as a consequence of the wave equation.

At $\Scri^+$ the relation $Q=0$ holds, and hence the wave equation implies the condition
\begin{equation}\label{BCScri}
  f_{,R}-f_{,T}=0.
\end{equation}
Similarly, at $\Scri^-$ one finds that $f_{,R}+f_{,T}=0$. Thus the tangential derivative of
$f$ along $\Scri$ vanishes, i.e. $f_{,\mu}t^\mu=0$, where
$t^\mu=(1, t^\vartheta, t^\varphi, \mp 1)$
is a tangential vector on $\Scri$. As a consequence, one may choose either the boundary condition 
\eqref{BCScri} at $\Scri^+$ or the Dirichlet condition
\begin{equation}\label{BCScri1}
  f=\textrm{constant}.
\end{equation}
In the latter case, the constant needs to be read off from the initial data which also extend up to $\Scri$.

\section{Numerical Method}\label{sec_num}

In this section we describe the numerical procedure for solving the
wave equation \eqref{wave_eq1} with the boundary conditions
\eqref{BCCenter} and \eqref{BCScri} or \eqref{BCScri1} for the two types
of mathematical problems --- the hyperboloidal IBVP and the Cauchy problem
(see Fig.~\ref{fig_IBVP}).

The numerical method consists of the following ingredients:
\begin{enumerate}
 \item Mapping of the physical domain onto a unit square (or onto several
       unit squares) by introducing appropriate coordinate transformations.
       The coordinates being defined on the unit square(s) are referred to as \emph{spectral coordinates}.
 \item Expressing the \lq\lq wave function\rq\rq\ $f$ in terms of another
       unknown function such that the initial conditions
       are satisfied automatically.
 \item Expansion of this unknown function in terms of a truncated series of \emph{Chebyshev
       polynomials} with respect to the spectral coordinates. A particular finite 
       resolution for the numerical approximation is chosen and 
       appropriate grid points in the spectral coordinates on the unit square(s) are identified.
 \item Formulation of an \emph{algebraic system of equations}
       for the values of the unknown function at the coordinate grid points. This system results 
       from the evaluation of the wave equation and 
       the boundary conditions within the spectral approximation scheme being chosen.
 \item Calculation of the solution of this system by means of the
       \emph{Newton-Raphson method}.
\end{enumerate} 

These points are described in some detail in the following
subsections.

\subsection{Spectral coordinates}\label{Sec_coordinates}

Smooth functions can be expressed in terms of spectral Chebyshev expansions provided
these functions are defined on an interval or, for more-dimensional functions,
on a cross product of intervals. For this reason we introduce an appropriate coordinate mapping through which 
the physical domain is obtained as the image of the unit square (in the 1+1-dimensional problem considered here). 

\begin{figure}[ht]
  \centerline{\psfig{file=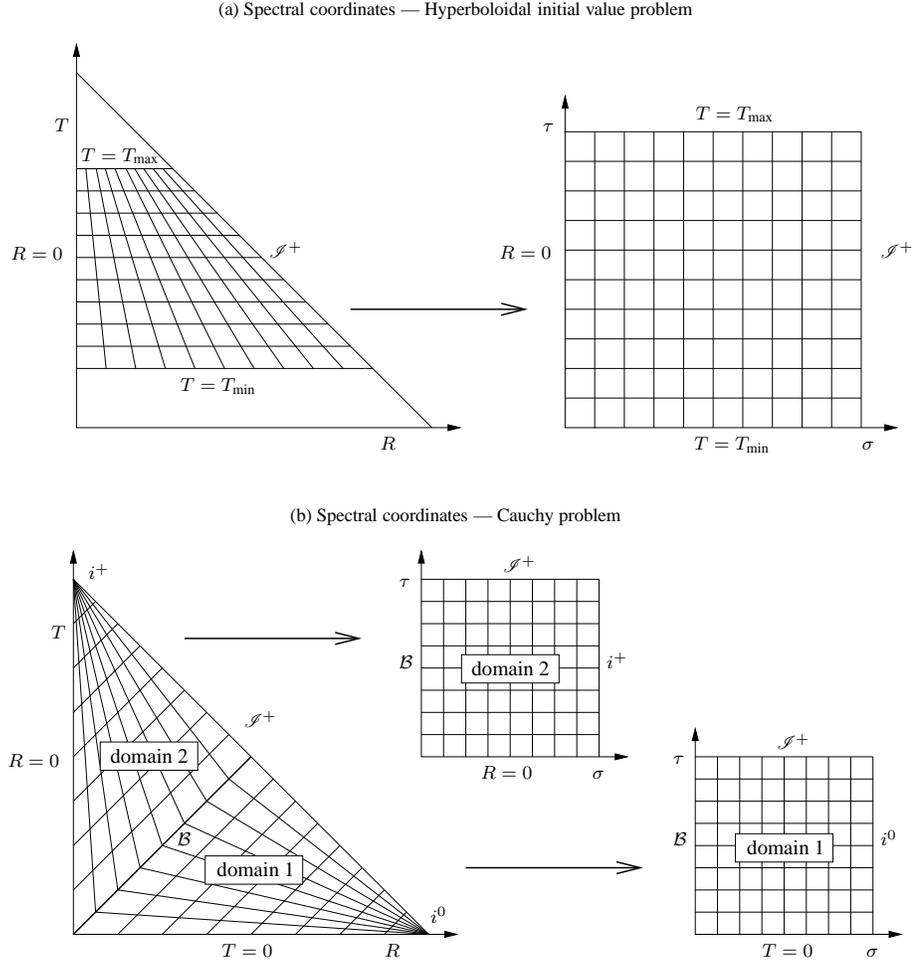,width=12cm}} 
  \vspace*{8pt}
  \caption{Illustration of the spectral coordinates $\sigma$ and $\tau$
  for the two types of IBVPs.
  \label{fig_coord}} 
\end{figure}

For the hyperboloidal IBVP we use the coordinate transformation
\begin{equation}
  R=\left[\frac{\pi}{2}-T_{\textrm{min}}
           -(T_{\textrm{max}}-T_{\textrm{min}})\tau\right]\sigma,\quad
  T=T_{\textrm{min}}+(T_{\textrm{max}}-T_{\textrm{min}})\tau,
\end{equation}
with $\sigma,\tau\in[0,1]$. An illustration of these coordinates
is given in Fig.~\ref{fig_coord}a, where particular
$\sigma$- and $\tau$-coordinate lines are shown. The physical boundaries
$T=T_\textrm{min}$, $T=T_\textrm{max}$, $R=0$ and $\Scri^+$ are mapped
to the edges $\tau=0$, $\tau=1$, $\sigma=0$ and $\sigma=1$ of the
square, respectively.

Note that at the points $i^0$ and $i^+$  the wave function is not analytic 
with respect to the coordinates $R$ and $T$. Therefore, in the case of the 
standard Cauchy problem, it is necessary to introduce a singular coordinate 
mapping through which the wave function becomes analytic with respect 
to the spectral coordinates.
To this end we divide up the physical domain into two triangular
subdomains and map 
each of these subdomains onto a unit square, see
Fig.~\ref{fig_coord}b. We use the particular
coordinates 
\begin{equation}
  R=\frac{\pi}{4}\left[2\sigma+(1-\sigma)\tau\right], \quad
  T=\frac{\pi}{4}(1-\sigma)\tau
\end{equation}
in domain 1 and
\begin{equation}
  R=\frac{\pi}{4}(1-\sigma)\tau, \quad
  T=\frac{\pi}{4}\left[2\sigma+(1-\sigma)\tau\right]
\end{equation}
in domain 2. The edges of the two squares correspond to $R=0$,
$T=0$, $i^0$, $i^+$, sections of $\Scri^+$, and, additionally, the
common boundary 
$\mathcal B=\{(R,T): R=T; 0\leq R \leq\pi/4\}$ between the two
domains. Boundary conditions at $\mathcal B$ can be derived from 
the requirement that the wave function $f$
be analytic in a neighbourhood of the initial slice $T=0$. 
As a consequence of the wave equation, $f$ is then analytic 
everywhere in the domain of dependence. In particular, for the 
standard Cauchy problem it follows that $f$ possesses a continuous and differentiable 
transition at $\mathcal B$.

The particular choice of the spectral coordinates is
not unique. Also the boundary $\mathcal B$ (in the Cauchy problem) 
could have been chosen to be at a different location (e.g. at
$T=\alpha R$ with $\alpha=\textrm{constant}\neq 1$).
Although our choice ($T=R$) represents a {\em characteristic} curve of the 
wave equation, in general the curve does not need to possess
specific features with respect to the underlying equation.

\subsection{Initial conditions}

A prescribed set of initial conditions for the wave equation, to be imposed at $T=T_\textrm{min}$ 
(i.e. at $\tau=0$), can be satisfied automatically through a specific ansatz for the wave function $f$.
For given functions $G(R)$ and $H(R)$ with
\begin{equation}
  f(R,T=T_{\textrm{min}})=G(R),\quad
  f_{,T}(R,T=T_{\textrm{min}})=H(R),
\end{equation}
we replace $f$, expressed as a function of the spectral coordinates
$\sigma$ and $\tau$, by a function $f_2$ via
\begin{equation}\label{Ansatz}
  f(\sigma,\tau)=f_0(\sigma)+f_1(\sigma)\tau+f_2(\sigma,\tau)\tau^2.
\end{equation}
The functions $f_0(\sigma)$ and $f_1(\sigma)$ are given in terms of the
initial data by
\begin{equation}\label{f0}
  f_0(\sigma)=f(\sigma,\tau=0)=G[R(\sigma,\tau=0)]
\end{equation}
and
\begin{equation}\label{f1}
  f_1(\sigma)=\diff{f(\sigma,\tau)}{\tau}\Big|_{\tau=0}
         = \diff{f}{R}\diff{R}{\tau}
           +\diff{f}{T}\diff{T}{\tau}\Big|_{\tau=0}
         = \diff{G}{R}\diff{R}{\tau}+H\diff{T}{\tau}\Big|_{\tau=0}.
\end{equation}

In this manner, for any regular choice of the new unknown quantity
$f_2(\sigma,\tau)$, the function
$f(\sigma,\tau)$ as given in \eqref{Ansatz} satisfies the correct
initial conditions at $\tau=0$.

\subsection{Chebyshev approximation}

In a spectral method the unknown functions are approximated in terms of 
appropriate basis functions. Here we choose \emph{Chebyshev polynomials} 
defined on the interval $[0,1]$. In particular, we approximate $f_2$ by
\begin{equation}\label{approx}
  f_2(\sigma,\tau) \approx \sum\limits_{i=0}^{N_\sigma}
         \sum\limits_{j=0}^{N_\tau}c_{ij}T_i(2\sigma-1)T_j(2\tau-1),
\end{equation}
where $T_n$ denotes the $n$th Chebyshev polynomial, $c_{ij}$ the
\emph{Chebyshev coefficients}, and
\begin{equation}
  n_\sigma=N_\sigma+1,\quad n_\tau=N_\tau+1
\end{equation}
are the prescribed resolution orders with respect to the spectral directions.

In accordance with this choice we introduce the following spectral \emph{collocation points} 
$P_{ij}=(\sigma_i,\tau_j)$
at which the the wave equation and the boundary conditions are evaluated in order to build up
an algebraic system of equations (see next subsection):
\begin{eqnarray}
  \sigma_i &=&
    \sin^2\!\left(\frac{\pi}{2}\frac{i}{N_\sigma}\right),\quad
    i=0,1,\dots,N_\sigma \\
  \tau_j  &=&  \sin^2\left(\frac{\pi}{2}\frac{j}{N_\tau}\right),\quad
    j=0,1,\dots,N_\tau.
\end{eqnarray}
Note that we choose to use the extrema of the Chebyshev polynomials
(Gauss-Lobatto collocation points)
so as to have gridpoints lying on the boundaries.

\subsection{Algebraic system of equations}\label{sec_alg}

For a given spectral approximation order, it is straightforward to
compute from the values $X_{ij}:=f_2(\sigma_i,\tau_j)$
\begin{enumerate}
	\item the Chebyshev coefficients of $f_2$
	\item the Chebyshev coefficients of the first and second derivatives of $f_2$ with respect 
		to the spectral coordinates $\sigma$ and $\tau$
	\item the values of these derivatives at the spectral collocation points $P_{ij}=(\sigma_i,\tau_j)$.
\end{enumerate}
For any values $X_{ij}$ it thus becomes possible to
\lq\lq evaluate\rq\rq\ the wave equation at the spectral gridpoints by inserting the function and derivative values. The mathematical task to be solved can therefore be
formulated as follows:
\emph{Calculate the $n_\sigma\times n_\tau$ unknown values $X_{ij}$ as
the  solution of the algebraic system of
the  $n_\sigma\times n_\tau$ equations $F_{ij}(X_{kl})=0$, where
$F_{ij}$ is the left hand side of 
the wave equation or boundary condition evaluated from $X_{kl}$ at the
grid-point $(i,j)$.} 

In the case of the hyperboloidal IBVP the following equations arise:
\begin{eqnarray}\label{cond1}
  &\textrm{\it at $R=0$}: f_{,R} = 0,\quad
  \textrm{\it at $\Scri^+$}: f_{,R}-f_{,T} = 0,\nonumber\\
  &\textrm{\emph{otherwise:} the wave equation},
\end{eqnarray}
and in the Cauchy problem
 \begin{eqnarray}\label{cond2}
  &\textrm{\it at $R=0$}: f_{,R} = 0,\quad
  \textrm{\it at $\Scri^+$, $i_0$ and $i^+$}: f = 0,\nonumber\\
  &\textrm{\emph{at $\mathcal B$:} $f$ and $\diff{f}{n}$ continuous},\quad
  \textrm{\emph{otherwise:} the wave equation},
\end{eqnarray}
where $\partial f/\partial n$ denotes the normal derivative
with respect to the boundary $\mathcal B$ (here,
$\partial f/\partial n\propto f_{,R}-f_{,T}$). The condition $f=0$ at
infinity is true for all initial data vanishing at infinity.
($f=0$ at infinity is automatically guaranteed for all
bounded functions $A(x)$
in the general solution \eqref{sol2} of the wave equation.)

It turns out that there are two exceptional points at which
the conditions \eqref{cond1} or \eqref{cond2} are already 
satisfied as a consequence of the \lq\lq ansatz\rq\rq\ \eqref{Ansatz}:
\begin{equation}\nonumber
	\mbox{Point 1:\quad} \sigma=\tau=0, \qquad\quad
        \mbox{Point 2:\quad} \sigma=1, \tau=0.
\end{equation}
These points correspond to the intersections of the initial slice $T=T_\textrm{min}$ with
$R=0$ and with $\Scri^+$ (in the hyperboloidal IBVP) or with $i^0$
(in domain 1 in the Cauchy problem). For example, at $\sigma=\tau=0$, the
condition $f_{,R}=0$ reduces to $g_{,R}=0$ which is already satisfied
for the spherically symmetric initial data.

Therefore, additional conditions need to be
imposed at these two points, in order to complete the algebraic system of equations.
As with the various other boundary conditions, these conditions also follow
from the wave 
equation \eqref{wave_eq1}. 

In the limit $R\to 0$ one obtains
\begin{equation}\label{ex1}
  3f_{,RR}-f_{,TT}-4\tan\!T\, f_{,T}=0.
\end{equation}

At $\Scri^+$ the following condition can be derived
\begin{equation}\label{ex2}
  f_{,RR}+f_{,TT}-2f_{,RT}=0.
\end{equation}

Finally, in the case of the Cauchy problem, at $\sigma=1$ and $\tau=0$
we impose the condition
\begin{equation}
  f_2 = 0.
\end{equation}
This is a consequence of \eqref{Ansatz} and the fact that $f=0$ at
$\Scri$. Since  
$f_0=f_1=0$ for $\sigma=1$, it follows that $f_2(\sigma=1,\tau\neq 0)=0$. 
Hence, for a continuous function, $f_2(\sigma=1,\tau = 0)=0$ holds.

\subsection{Newton Raphson method}

Throughout  most of this paper we concentrate on the linear wave
equation, for which the system $F_{ij}(X_{kl})=0$ is linear. A
corresponding solution method as e.g. the Gauss-Jordan elimination or
the LU decomposition would provide us with the solution to our
IBVP. However, in order to tackle non-linear equations it is
necessary to use a more general scheme. For this reason 
we choose the Newton Raphson method.   
 
We define the $n_\sigma\times n_\tau$-dimensional vectors
\begin{eqnarray}
  \mathbf X &:=& (X_{00}, X_{01},\dots,X_{0N_\tau},
   X_{10},\dots,X_{1N_\tau},\dots, X_{N_\sigma 0}, \dots,
   X_{N_\sigma N_\tau}),\quad\\
  \mathbf F &:=& (F_{00}, F_{01}\dots,F_{0\,N_\tau},
   F_{10},\dots,F_{1N_\tau},\dots,F_{N_\sigma\, 0},
   \dots,F_{N_\sigma N_\tau}),
\end{eqnarray}
containing all components of $X_{ij}$ and $F_{ij}$. Hence, the
system of equations can be written as
\begin{equation}\label{alg_eq}
\mathbf F(\mathbf X)= {\bf 0}. 
\end{equation}
Within the Newton Raphson method, the system \eqref{alg_eq} is solved
iteratively using an \lq\lq initial guess\rq\rq\ $\mathbf X^0$
and computing the successive vectors
\begin{equation}
  \mathbf X^{n+1}=\mathbf X^n
    - [\mathbf F'(\mathbf X^n)]^{-1}\mathbf F(\mathbf X^n).
\end{equation}
Here we approximate the \emph{Jacobi matrix} $\mathbf F'=(\partial F_i/\partial X_j)$
through a second order finite difference representation and obtain the inverse
via an LU decomposition\footnote{Note that for our 1+1-dimensional IBVP
a direct matrix inversion is computationally feasible. However, for
higher dimensional problems
such methods are too computationally expensive and one requires the use
of iterative matrix inversion methods.}. 

\section{Test of the Method with Explicit Examples}\label{sec_examples}
\subsection{Numerical accuracy}

We study several explicit examples in order to investigate the
effectiveness of our numerical method. 
From \eqref{sol2} we can construct a particular solution to the wave equation by choosing the appropriate function $A(x)$. 
For a given time interval $[T_\textrm{min},T_\textrm{max}]$,
where $0\le T_\textrm{min} < T_\textrm{max}\le \frac{\pi}{2}$ (equal
signs in the case of the standard Cauchy problem), we read off $f$ and $f_{,T}$ at $T=T_\textrm{min}$ as initial data for the
numerical method and solve for $f$ in the entire domain expanding up to the maximal time $T=T_\textrm{max}$.

As a measure of the overall numerical accuracy, we calculate the
\emph{numerical residual}
\begin{equation}\label{Rnum}
R_{\textrm{num}}(n_\sigma, n_\tau):=
          \max\limits_{(\sigma,\tau)\in[0,1]^2}
          \left|f_{\textrm{num}}(\sigma,\tau)
               -f_{\textrm{an}}(\sigma,\tau)\right|,
\end{equation}
where $f_\textrm{num}$ and $f_\textrm{an}$ denote the
numerical and analytical solution, respectively.

In order to investigate to what extent the numerical method is effective
and leads to accurate 
solutions, we compare $R_\textrm{num}$ to the \emph{analytical} error caused by an exact Chebyshev representation of the given resolution order. This error is given by the following \emph{analytical residual}
\begin{equation}\label{Rana}
  R_{\textrm{an}}(n_\sigma, n_\tau):=
        \max\limits_{(\sigma,\tau)\in[0,1]^2}
        \left|f_{\textrm{approx}}(\sigma,\tau)
        -f_{\textrm{an}}(\sigma,\tau)\right|
\end{equation}
and describes the maximal difference between the
analytical solution $f_\textrm{an}$ and its Chebyshev
approximation
\begin{equation}\label{Chebyshev_app_2D}
  f_{\textrm{approx}}
         =\sum\limits_{i=0}^{n_\sigma-1}\sum\limits_{j=0}^{n_\tau-1}
          \tilde c_{ij}T_i(2\sigma-1)T_j(2\tau-1)
\end{equation}
of the order ($n_\sigma$, $n_\tau$)\footnote{For the calculation of the coefficients $\tilde c_{ij}$ we have used the {\em exact} values of the function $f_\textrm{an}$ at the gridpoints $(\sigma_i,\tau_j)$.}.

Since the error \eqref{Rana} of the Chebyshev approximation \eqref{Chebyshev_app_2D} is close to the smallest
possible polynomial approximation error
(which one would obtain for the \emph{optimal} approximation
polynom of the particular function to be approximated),
$R_\textrm{an}$ provides a good measure for
the effectiveness and potential accuracy of our numerical method. 
In the subsequent subsection, we display plots of both residuals $R_\textrm{num}$ and $R_\textrm{an}$ 
for several example solutions.

A well-known property of Chebyshev approximations for smooth
functions is an exponential decay of $R_\textrm{an}$ with increasing
resolution (up to a final saturation level of order $10^{-14}$ or
$10^{-15}$ due to the finite machine accuracy of $16$ numerical digits being used).
As will be demonstrated below, the numerical residuals $R_\textrm{num}$
also possess an exponential fall-off and reach a final saturation level near
the final level of $R_\textrm{an}$.

Another important issue is the dependence of the numerical residual
on the size of the time interval
$[T_\textrm{min},T_\textrm{max}]$.
It turns out that without substantial loss of accuracy we may choose a
time step that is comparable to or larger than
a spatial scale implied by the initial data.

\subsection{Numerical examples}

As a first example we consider $A(x)=x$ in Eq. \eqref{sol2},
i.e. the analytical solution
\begin{equation}
  f(R,T)=\frac{2R}{\tan(R+T)+\tan(R-T)}.
\end{equation}

\begin{figure}[ht]
  \centerline{\psfig{file=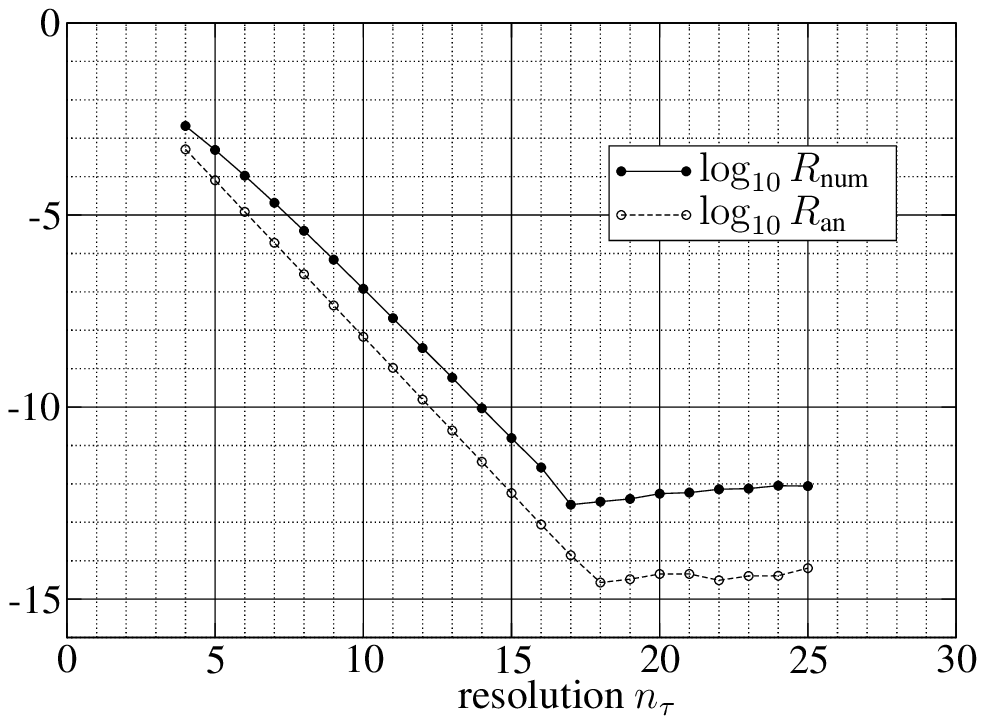,width=6.1cm}\hfill
              \psfig{file=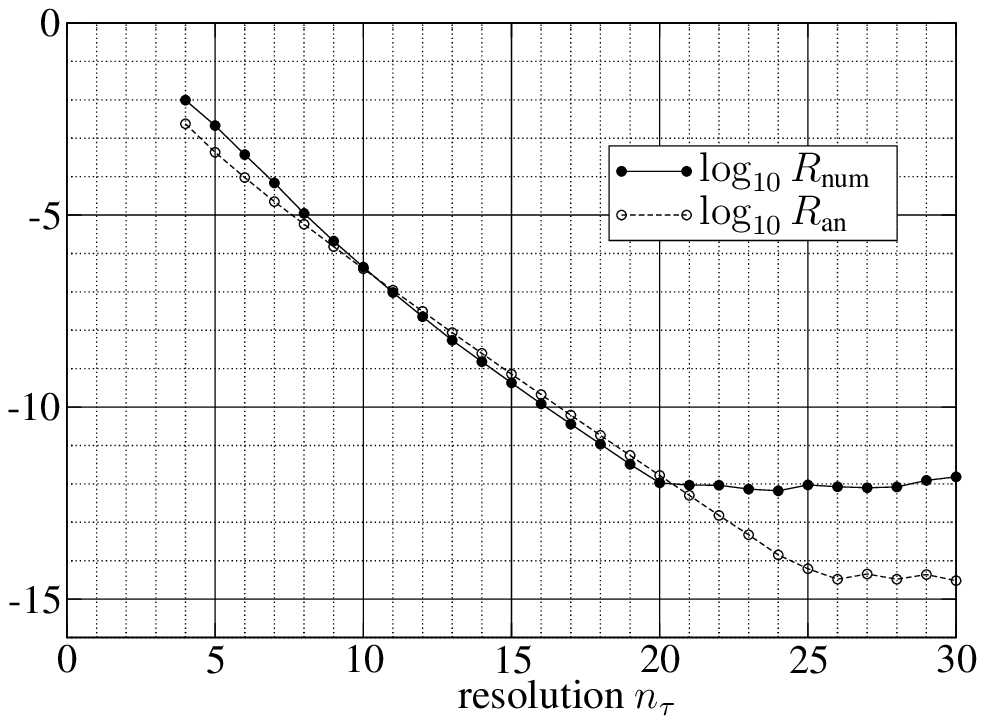,width=6.1cm}
   } 
  \vspace*{8pt}
  \caption{Plot of the numerical and analytical
  residuals for the example solution with $A(x)=x$.
  Parameters: $n_\sigma=2n_\tau$,
  $T_\textrm{min}=0.3$. \emph{Left panel:} $T_\textrm{max}=0.6$.
  \emph{Right panel:}  $T_\textrm{max}=1.2$.
  (Note that with the latter value of
  $T_\textrm{max}$ a fairly large time interval
  is realized since future time-like infinity is characterized by
  $T=\pi/2=1.57\dots$)
  \label{Sol0}} 
\end{figure}

(A plot of this solution can be found in Fig.~\ref{Plot1} in
\ref{Appendix}.)
The numerical and analytical residuals [as defined in \eqref{Rnum},
\eqref{Rana}]  for a small and a large time
interval in the hyperboloidal IBVP are shown in Fig.~\ref{Sol0}.
We have chosen the resolutions $n_\sigma$ and $n_\tau$ to be connected
by $n_\sigma=2n_\tau$.

As expected, $R_\textrm{an}$ exhibits \emph{geometric convergence}
(the graph is roughly a straight line in the logarithmic plot).
For $n_\tau=18$ a saturation level is reached at a value below $10^{-14}$. 
The numerical residual $R_\textrm{num}$ shows the same
qualitative behaviour: It decreases almost linearly in the logarithmic
plot until it reaches a saturation level at a value of about $10^{-12}$
for $n_\tau=17$ in the case of the small time interval and also at almost
$10^{-12}$ for $n_\tau=20$ in the case of the large time interval. 
Thus approximately the same accuracy of about $12$ correct digits 
can be obtained for both the small and the large time interval.

\begin{figure}[ht]
  \centerline{\psfig{file=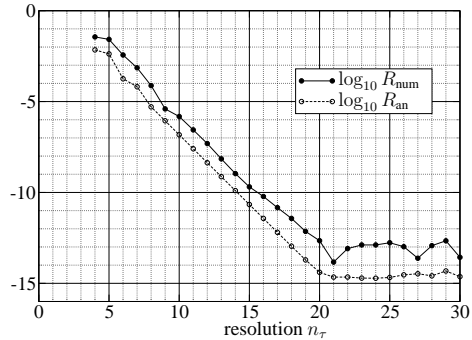,width=6.1cm}} 
  \vspace*{8pt}
  \caption{Numerical accuracy in the standard Cauchy problem with $A(x)=x$.
  The numerical resolutions are chosen to be related by $n_\sigma=n_\tau$
  in both subdomains. 
  \label{Sol0Cauchy}} 
\end{figure}

We study the solution in question with $A(x)=x$ also as a
standard Cauchy problem. The numerical result is presented in
Fig.~\ref{Sol0Cauchy}. The final saturation level
of about $10^{-13}$ (which is reached for $20\times20$ collocation points
in each numerical domain) is even below the final level in the hyperboloidal
problem. The reason for this might be the particular treatment of the points $i^0$ and $i^+$, see Sec.~\ref{Sec_coordinates}.

\begin{figure}[ht]
  \centerline{\psfig{file=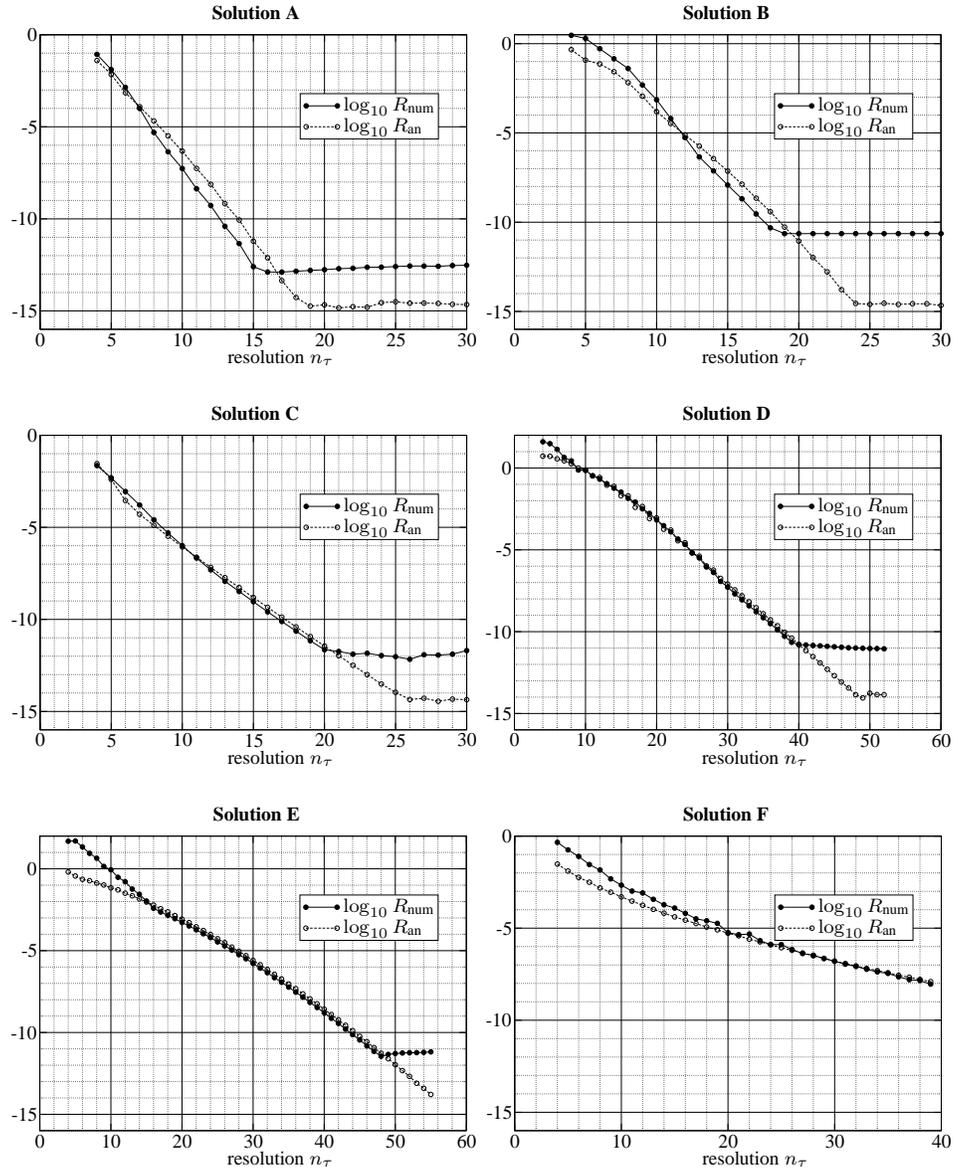,width=12.5cm}}
  \vspace*{8pt}
  \caption{The residuals of the examples for the hyperboloidal IBVP in
  Tab.~\ref{tab}, see Fig.~\ref{PlotsA-F}. 
  \label{SolAF}} 
\end{figure}

A number of further numerical examples for the hyperboloidal IBVP
are summarized in
Tab.~\ref{tab}.
(For plots of these solutions see Fig.~\ref{PlotsA-F} in \ref{Appendix}.)
The corresponding plots of the numerical accuracies are
shown in Fig.~\ref{SolAF}. In these examples,
a saturation level between
$10^{-10}$ and $10^{-13}$ is reached for sufficiently large spectral
resolutions. An exception is example F, which describes an initial pulse on a compact support.
Even with a resolution of
$n_\sigma=160$, $n_\tau=40$ (where the saturation level is not yet
reached) the numerical solution is correct only in the first $8$ digits. This follows from the fact that
a very accurate spectral approximation of a non-analytic $C^\infty$-function requires high resolution. As a consequence, the numerical techniques become expensive.

\begin{table}[ht]\centering
\caption{List of the further numerical examples for the hyperboloidal
IBVP, see Figs.~\ref{SolAF} and \ref{PlotsA-F}.
 \label{tab}}
  \begin{tabular}{cccccc}
   \toprule
     & $A(x)$ & Remark  & $T_\textrm{min}$ & $T_\textrm{max}$ &
      $n_\sigma$\\
   \colrule
     (A) &
     $\frac{1}{b}\sin(bx)$ with $b=10$ &
     \begin{minipage}[t]{3.5cm}
     arbitrary number of\\
     minima and maxima\\
     depending on $b$
     \end{minipage} &
     $0.3$ & $0.6$ & $2 n_\tau$\\
  \colrule
     (B) &
     $\frac{1}{b}\sin(bx)$ with $b=20$ &
      cf. (A) &
     $0.3$ & $0.6$ & $2 n_\tau$\\
   \colrule
     (C) &
     $(x-T_\textrm{min})^2(x-T_\textrm{min}+1)$  &
     \begin{minipage}[t]{3.5cm}
     incoming \lq\lq hill\rq\rq
     \end{minipage} &
     $0.3$ & $1.2$ & $2 n_\tau$\\
   \colrule
     (D) &
     $\ee^{-64(x-0.7)^2}$ &
     \begin{minipage}[t]{3.5cm}
     incoming Gauss-like\\ pulse, reflected at\\ $R=0$, outgoing with\\
     inversed amplitude
     \end{minipage} &
     $0.3$ & $1.0$ & $2 n_\tau$\\
   \colrule
     (E) &
     $\ee^{-64(x-1.1)^2}-\ee^{-64(x+0.1)^2}$ &
     \begin{minipage}[t]{3.5cm}
     two Gauss-like pulses\\ crossing each other
     \end{minipage} &
     $0.3$ & $0.9$ & $2 n_\tau$\\
   \colrule
     (F) &
     \begin{minipage}[t]{3.3cm}
     $-30\ee^{-\frac{1}{(x+0.8)(0.2-x)}}$\\
     (if $-0.8<x<0.2$,\\
     otherwise $A(x)=0$)
     \end{minipage} &
     \begin{minipage}[t]{3.5cm}
     outgoing pulse with\\ compact support\\
     ($C^\infty$, not analytic)
     \end{minipage} &
     $0.3$ & $0.6$ & $4 n_\tau$\\
    \botrule
  \end{tabular}
\end{table}

\section{Other Equations}\label{sec_other}

The numerical examples presented in the previous section provide evidence for the fact that
highly accurate solutions to the wave equation can be obtained using a
fully pseudospectral scheme.  
In this section we demonstrate that the method also works in the case of other differential equations. 
We discuss two particular modifications of the wave equation:
an inhomogeneous wave equation and a non-linear wave equation.  

\subsection{Inhomogeneous wave equation}

An interesting feature of time evolution problems
in General Relativity is the possible formation of singularities from
completely regular initial data, 
e.g. the scenario of the formation of a black hole from a collapsing star.
For this reason we investigate the applicability of our numerical method
to an example 
solution which develops a pole like singularity. In particular we study
how closely to this critical point the numerical 
domain can be located. Since solutions to the homogeneous linear wave
equation do not develop 
singularities from regular initial data, we consider a particular
\emph{inhomogeneous} linear wave equation, 
\begin{equation}\label{inh_eq}
   QP(f_{,RR}-f_{,TT})
  +2\frac{(Q^2+P^2)f_{,R}+(Q^2-P^2)f_{,T}}{\sin(2R)}=I(R,T),
\end{equation}
where the right hand side $I(R,T)$ is taken such that $f$
develops a pole after a finite time. This property is guaranteed by choosing the solution 
\begin{equation}\label{inh_sol}
  f(R,T) = \frac{1}{(R^2-R_0^2)^2+(T^2-T_0^2)^2},
\end{equation}
which is singular for $R=R_0$, $T=T_0$. From this expression
we can calculate the
corresponding inhomogeneity $I$ as well as the initial data. Equation
\eqref{inh_eq} can be solved in the same manner as the homogeneous wave equation. 
Note that the boundary conditions at $\Scri^+$ and at the \lq\lq exeptional points\rq\rq, cf.  
Sec.~\ref{sec_alg}, take on a different form. At $\Scri^+$, instead of \eqref{BCScri} we impose the condition 
\begin{equation}
  4\sin(2R)(f_{,R}-f_{,T})=I.
\end{equation}
For $\sigma=\tau=0$ we rewrite \eqref{ex1} accordingly:
\begin{equation}
  3f_{,RR}-f_{,TT}-4\tan T f_{,T}=\frac{I}{2\cos^2T}.
\end{equation}
Finally, at $\sigma=1$, $\tau=0$, we replace \eqref{ex2} by the condition
\begin{equation}
  2\sin(2R)(f_{,RR}+f_{,TT}-2f_{,RT})=I_{,R}.
\end{equation}

\begin{figure}[ht]
  \centerline{\psfig{file=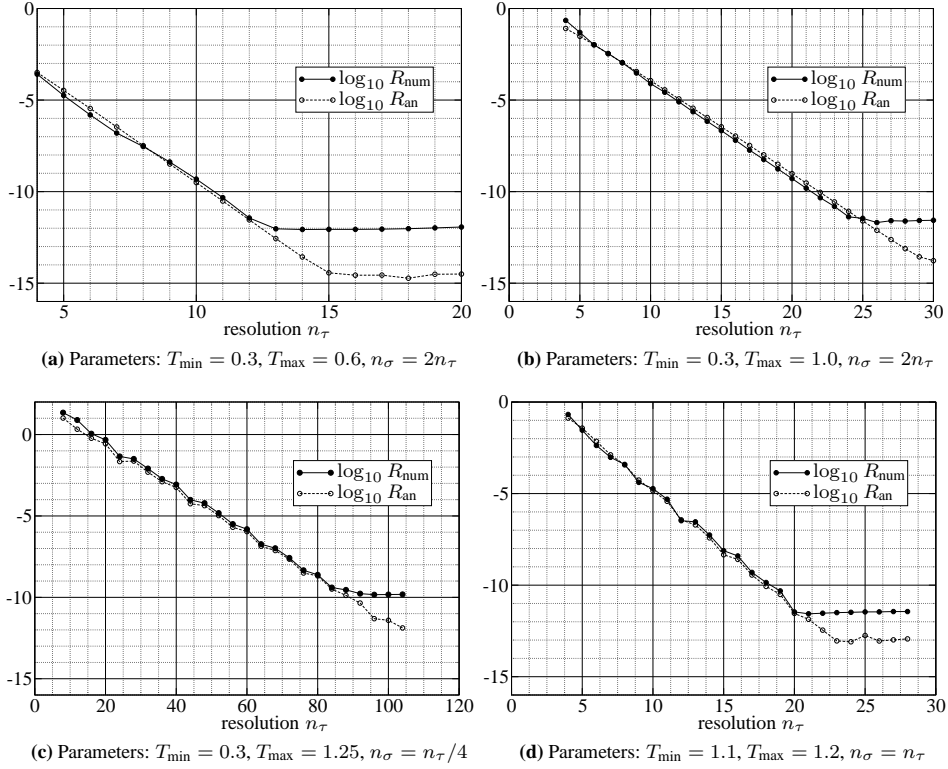,width=12.5cm}}
  \vspace*{8pt}
  \caption{The residuals $R_\textrm{num}$ and $R_\textrm{an}$ for solutions of the inhomogeneous wave equation, 
  being calculated on different time intervals $[T_\textrm{min},T_\textrm{max}]$. The
  pole is located at $R_0=0.1$, $T_0=1.3$. A plot of the solution is
  shown in Fig.~\ref{PlotInh} in \ref{Appendix}.
  \label{SolInh}} 
\end{figure}

The residuals for the numerical solution of the hyperboloidal IBVP with
the exact solution \eqref{inh_sol} are shown in
Fig.~\ref{SolInh}. In \ref{SolInh}a-c, the solution is calculated in a
hyperboloidal domain with a fixed value of $T_\textrm{min}=0.3$ and
increasing maximal time values $T_\textrm{max}$ approaching the
singularity. One observes that more and more  resolution with respect to
the time 
direction is required in order to reach the saturation level. Furthermore, the
maximal accuracy obtained decreases slightly. However, even in a
small vicinity of the pole, 
the solutions are very accurate for a sufficiently large resolution.

In Fig.~\ref{SolInh}d, the inhomogeneous wave equation is solved within a
narrow hyperboloidal strip with $T_\textrm{min}=1.1$ and
$T_\textrm{max}=1.2$. Although the singularity (located at $R_0=0.1$, $T_0=1.3$) is close to this
domain, the solution is highly accurate (11 correct digits) and the
numerical saturation is reached for a moderate value of $n_\tau=20$.

The above examples demonstrate the applicability of our fully spectral method to solutions which encounter a singular behaviour. The numerical solutions being obtained retain high accuracy, provided that an appropriate resolution is chosen.

\subsection{Non-linear wave equation}\label{sec_NL}

In the preceding sections we have studied \emph{linear} differential
equations. However, in view of future applications in General
Relativity, it is interesting to apply our numerical method to
\emph{non-linear} equations. (In an appropriate formulation, the
Einstein equations reduce to a set of non-linear wave equations.)

To this end, we consider the example equation
\begin{equation}\label{nl_eq}
  \Box f \equiv f_{,rr}+\frac{2}{r}f_{,r}-f_{,tt}
         =\lambda\left[r(f_{,r}^2-f_{,t}^2)
         +2f{}f_{,r}+\frac{1}{r}f^2\right]
\end{equation}
with $\lambda=\textrm{constant}$. The general solution (regular at
$r=0$) is given by\footnote{The solution can be obtained by
writing $f$ as $f(r,t)=\Phi(r,t)/r$ and introducing null coordinates $u=r+t$,
$v=r-t$. Then, Eq. \eqref{nl_eq} reduces to
$\Phi_{,uv}=\lambda\Phi_{,u}\Phi_{,v}$, which can be solved easily.}
\begin{equation}
  f(r,t)=-\frac{1}{\lambda
       r}\ln\left[1+\lambda\left(a(t+r)-a(t-r)\right)\right].
\end{equation}

We introduce again the coordinates $R$ and $T$ [see \eqref{coord}] and
obtain the non-linear wave equation
\begin{align}\label{nl_eq1}
  &QP\left[QP(f_{,RR}-f_{,TT})+\frac{2}{\sin(2R)}
  \left[(Q^2+P^2)f_{,R}+(Q^2-P^2)f_{,T}\right]\right]\nonumber\\
 & =\lambda\left[\sin(2R)QP(f_{,R}^2-f_{,T}^2)
   +2f\left[(Q^2+P^2)f_{,R}+(Q^2-P^2)f_{,T}\right]
   +4\frac{QP}{\sin(2R)}f^2\right]
\end{align}
with $Q$ and $P$ as defined in \eqref{QP}.
The general solution reads
\begin{equation}\label{sol_nonlin}
  f(R,T)=-\frac{2}{\lambda}\frac{\ln\left[1
         +\lambda(A(T+R)-A(T-R))\right]}{\tan(R+T)+\tan(R-T)}.
\end{equation}
As with the linear wave equation, the boundary
conditions can be obtained by analyzing the equation at the singular
points. In this manner we find the conditions
\begin{equation}
  Q^2f_{,R}=\lambda f^2\quad \textrm{at}\quad R=0,\quad
  f_{,R}-f_{,T}=0\quad \textrm{at}\quad \Scri^+
\end{equation}
and, additionally, at the \lq\lq exceptional points\rq\rq\ (see
Sec.~\ref{sec_alg})
\begin{equation}
  3f_{,RR}-f_{,TT}-4\tan T f_{,T}=\frac{4\lambda}{\cos^2 T}f f_{,R}
  \quad \textrm{at}\quad  \sigma=0, \tau=0,
\end{equation}
and
\begin{equation}
  f_{,RR}+f_{,TT}-2f_{,RT}=0
   \quad \textrm{at}\quad \sigma=1, \tau=0.
\end{equation}
Note that in the limit $\lambda\to 0$ one recovers the boundary conditions of the \emph{linear} wave equation.

\begin{figure}[ht]
  \centerline{\psfig{file=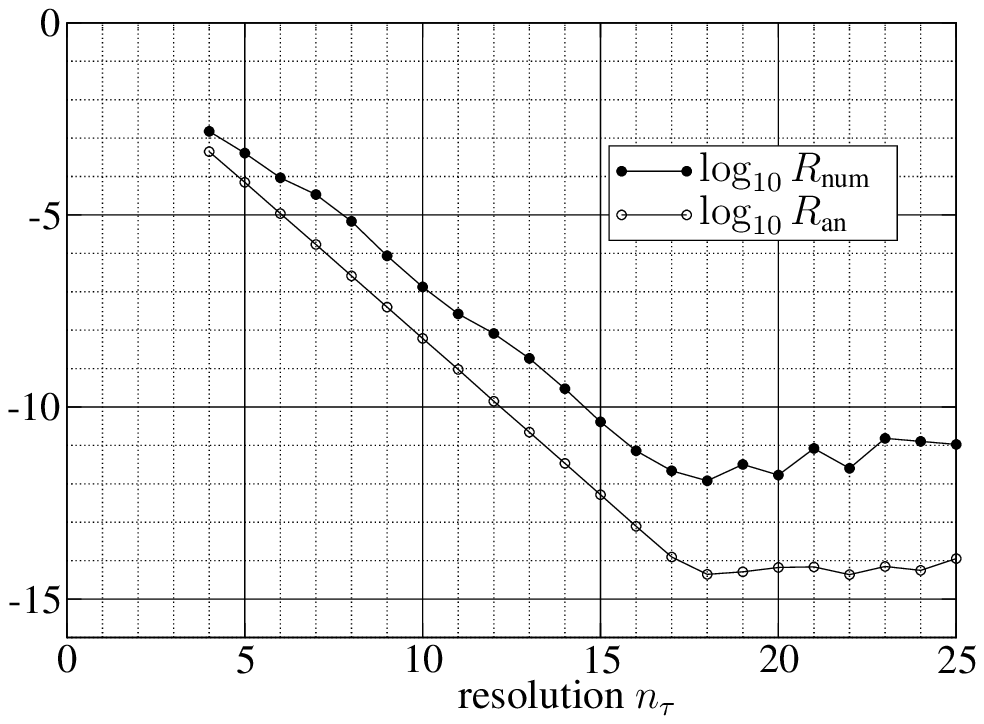,width=6.1cm}\hfill
              \psfig{file=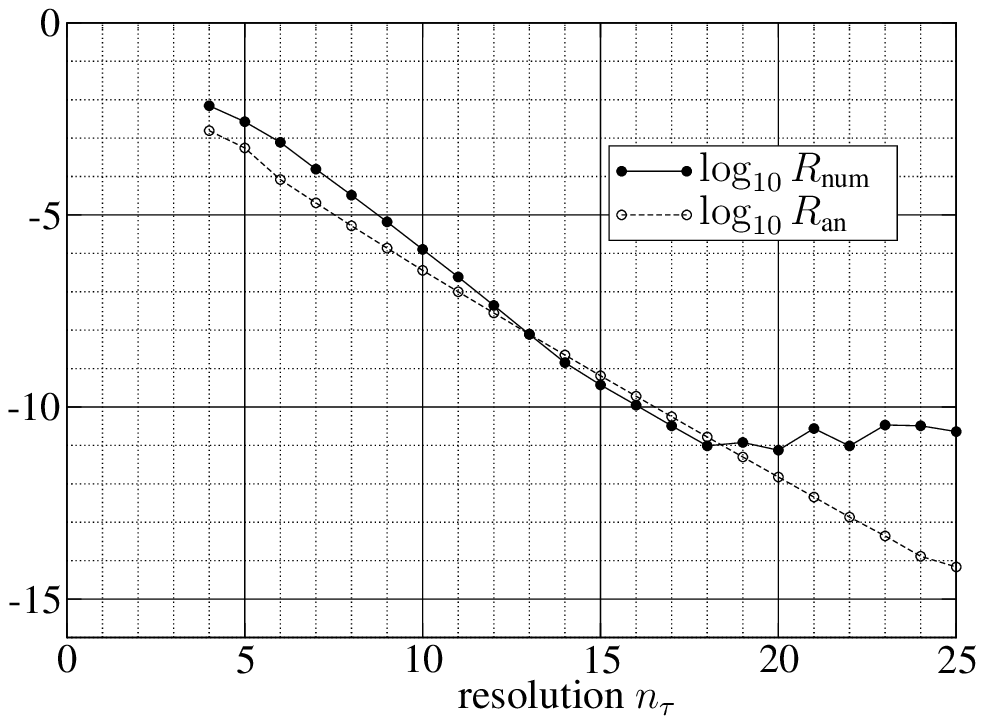,width=6.1cm}
   } 
  \vspace*{8pt}
  \caption{The numerical and analytical
  residuals for the example solution to the non-linear
  wave equation with $A(x)=x$, see \eqref{sol_nonlin}.
  Parameters: $\lambda=1$, $n_\sigma=2n_\tau$,
  $T_\textrm{min}=0.3$. \emph{Left panel:} $T_\textrm{max}=0.6$.
  \emph{Right panel:}  $T_\textrm{max}=1.2$, see Fig.~\ref{PlotNL} for a
  plot of the solution.
  \label{Solnonlin}} 
\end{figure}

As an example we choose the particular solution \eqref{sol_nonlin}
obtained for $A(x)=x$ (see Fig.~\ref{PlotNL} in \ref{Appendix}).  
Again, we read off the corresponding initial data and apply our numerical method to solve the IBVP. The results are
displayed in Fig.~\ref{Solnonlin}.

We have solved the hyperboloidal IBVP for both a small and  a large time
interval (left and right panel in Fig.~\ref{Solnonlin}). One finds 
a numerical accuracy of about $10$ to $11$ correct digits. Hence, the
solution is nearly as precise as the corresponding example solution
of the \emph{linear} wave equation with $A(x)=x$, cf. Fig.~\ref{Sol0}.

\section{Regularized wave equation}\label{sec_regularization}

An important point in the preceding discussion of the wave equation is the
degeneracy at $\Scri$, which is a consequence of the conformal
approach. This degeneracy provided us with a particular first order
boundary condition to be imposed at $\Scri$. As an alternative to the
numerical solution of this \emph{singular} 
equation, we investigate in this section a \emph{regularized}
version. The motivation behind this study  
is our future goal of solving the Einstein equations numerically on a
conformally compactified space-time. Note that a regular formulation of
Einstein's equations can be identified (Friedrich's regular conformal
field equations, \cite{Friedrich}). 

For the derivation of the regularized wave equation, consider 
two conformally related $n$-dimensional metrics $g$ and $\bar g$ with
\begin{equation}
  \bar g_{\mu\nu}=\Omega^2 g_{\mu\nu},
\end{equation}
where $\Omega=\Omega(x^\mu)$ is a conformal factor. Then, the relation
\begin{equation}
  \left(\bar\Box - \frac{n-2}{4(n-1)}\bar R\right)
    \left(\Omega^{1-\frac{n}{2}} f\right)
    = \Omega^{-1-\frac{n}{2}}\left(\Box - \frac{n-2}{4(n-1)} R\right) f
\end{equation}
holds, see \cite{Wald}.
$\bar R$ and $R$ denote the Ricci scalars of the two conformal
metrics, and $\bar\Box$ and $\Box$ are the wave
operators\footnote{
The wave operators are defined by
$\bar\Box = \bar g^{\mu\nu}\bar\nabla_{\mu}\bar\nabla_{\nu}$
and
$\Box = g^{\mu\nu}\nabla_{\mu}\nabla_{\nu}$.}.
In our case we choose $\Omega=PQ$ [cf.~\eqref{line}].
With $n=4$, $R=0$
(flat Minkowski space-time), $\bar R=6$,
and with the definition $\bar f := \Omega^{-1}f$
we obtain
\begin{equation}
  \Box f = \Box (\Omega\bar f)=\Omega^3(\bar\Box-1)\bar f.
\end{equation}
Hence, equivalently to the singular equation $\Box f=0$, we may study the
regular equation
\begin{equation}\label{reg}
  (\bar\Box-1)\bar f
  \equiv \frac{1}{4}(\bar f_{,RR}-\bar f_{,TT})+\frac{\bar f_{ ,R}}{\tan(2R)}
    -\bar f = 0.
\end{equation}

This equation does \emph{not} reduce to a first order
boundary condition at $\Scri^+$. For the completeness of the algebraic 
system discussed in section \ref{sec_alg} we therefore need to require
the nonlinear second 
order wave equation as a condition there.
(This second order boundary condition can be imposed since no
\lq\lq information from outside\rq\rq\ can encounter the numerical
domain, i.e. no ingoing characteristic crosses the boundary in question.)
 
Again we study an explicit example by looking at the hyperboloidal IBVP 
(C) in Tab.~\ref{tab}, but this time we solve the \emph{regular} wave equation
\eqref{reg}. It turns out that this formulation of the problem
also permits highly accurate numerical solutions. However, instead of $12$
correct numerical digits (as in Fig.~\ref{SolAF}C) we obtain $11$
digits, thereby loosing one 
order in the numerical accuracy.


Note that $\bar f$ possesses a similar accuracy as the normal derivative
$\partial f/\partial n$ with respect to ingoing characteristics. In
particular, the content of outgoing radiation at $\Scri^+$, described by
\begin{equation}
 \bar f|_{\Scri^+}=\lim\limits_{\Omega\to0}\frac{f}{\Omega}\propto
\diff{f}{n}\big|_{\Scri^+},
\end{equation}
is similarly precisely given by $\bar f$
(calculated from the regularized wave equation)
and by $\partial f/\partial n$
(calculated from the singular wave equation).

We conclude that, remarkably,
our spectral algorithm is applicable to both regular and degenerate wave equations. 
For the fully spectral scheme, the particular formulation of a given equation 
seems to play a subordinate role. 

\section{Discussion}\label{sec_discussion}

In this paper, we constructed numerical solutions of hyperbolic
equations utilizing a fully pseudospectral scheme. Combining this
method with the concept of conformal infinity, we were able to obtain
highly accurate solutions.

Interestingly, by means of the method presented,
hyperbolic equations can be handled quite
similarly to elliptic equations. There is no principal 
difference between the treatment of the
spatial coordinate $\sigma$ and that of the time coordinate $\tau$
in the Chebyshev approximation
\eqref{approx}.

The formulation of boundary/initial values is, however, 
fundamentally different for elliptic and hyperbolic problems.
In an elliptic problem, at each boundary \emph{one} first order condition 
is required, i.e. a condition which contains no second order normal 
derivative with respect to the boundary in question. 
On the other hand, a hyperbolic problem requires knowledge of 
the unknown function and its normal (time-) derivative at the initial slice, 
i.e. \emph{two} first order conditions there. The future boundary
$T=T_\textrm{max}$  
is treated like an interior point, i.e. the hyperbolic equation 
provides a second order condition there. Note that any attempt to impose
a first order condition at this boundary must fail\footnote{As an
example, for the wave equation  
a particular Dirichlet type boundary value problem can be shown to admit
infinitely many solutions.} --- not only for our pseudospectral algorithm
but for any numerical scheme. 

The method being presented possesses a highly implicit character. The time
evolution is not performed successively, 
moving from one time slice to the next, 
but the entire system is solved simultaneously instead. As a
consequence, it turned out that for all our numerical
examples it was not necessary to 
respect a Courant-Friedrichs-Lewy (CFL) condition. In fact,
the spatial collocation points could be distributed much more densely than
the time points.

We have demonstrated that for solutions which admit a rapidly converging
Chebyshev expansion,  
the method leads to an accuracy of up to $12$ or $13$ correct digits
(for a double precision code). 
In all examples being discussed, the numerical error ($R_\textrm{num}$)
is close to the analytical error 
($R_\textrm{an}$), i.e. the algorithm proves to be very
effective. Furthermore, a geometric convergence rate is exhibited, i.e.  
the error decreases exponentially with the resolution.

Our results encourage us to attempt to develop similar numerical
techniques to solve the dynamical Einstein equations. As an interesting
area of application we envision time evolution problems of perturbed
axisymmetric equilibrium configurations (as e.g. rotating stars, rings
or black holes with surrounding matter). 
Consequently, the method would permit a highly accurate stability
analysis of such objects and a careful investigation of the emitted
gravitational waves at $\Scri$.
 
In order to apply our pseudospectral scheme to such problems,
it will be necessary to cover the
space-time with more than a single computational domain. (At least
separate domains for matter and vacuum regions are required.) For the
treatment of non-spherically symmetric equations (i.e. higher
dimensional problems), the computationally expensive
direct matrix inversion in the Newton Raphson
method needs to be replaced by an iterative inversion method, as already
mentioned earlier.
We believe that through the implementation of appropriate technical
details our method becomes applicable to the solution of such physically
interesting problems.

\section*{Acknowledgments}

We would like to thank Helmut Friedrich, J\'er\^ome Novak, Silvano
Bonazzola, and David Petroff for many valuable
discussions. This work was supported by the Deutsche
Forschungsgemeinschaft (DFG) through the
Collaborative Research Centre SFB/TR7
\lq\lq Gravitational wave astronomy\rq\rq. 

\newpage

\appendix

\section{Plots of the example solutions}\label{Appendix}

Within this appendix we provide plots of all numerical example
solutions to the linear, inhomogeneous and non-linear wave equation
that we have studied in this paper.

\begin{figure}[ht]
  \centerline{\psfig{file=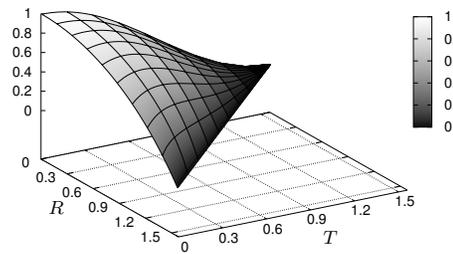,width=6cm}} 
  \vspace*{8pt}
  \caption{Solution to the linear wave equation
  with $A(x) = x$, cf. Figs. \ref{Sol0} and \ref{Sol0Cauchy}. 
  \label{Plot1}} 
\end{figure}

\begin{figure}[ht]
  \centerline{\psfig{file=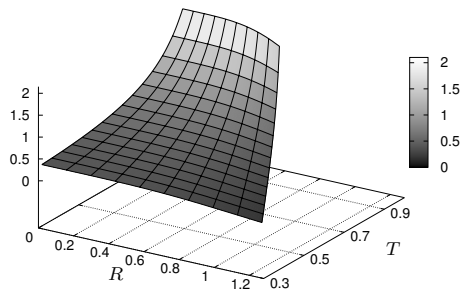,width=6cm}} 
  \vspace*{8pt}
  \caption{Example solution to the inhomogeneous
  wave equation, cf. Fig. \ref{SolInh}. 
  \label{PlotInh}} 
\end{figure}

\begin{figure}[ht]
  \centerline{\psfig{file=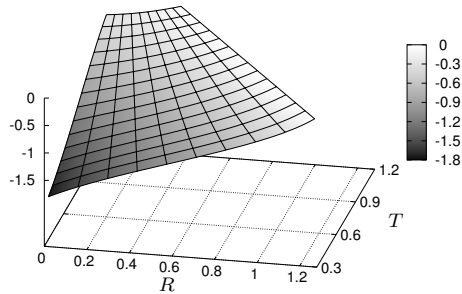,width=6cm}} 
  \vspace*{8pt}
  \caption{Example solution to the non-linear
  wave equation, cf. Fig. \ref{Solnonlin}. 
 \label{PlotNL}} 
\end{figure}

\begin{figure}[ht]
  \centerline{\psfig{file=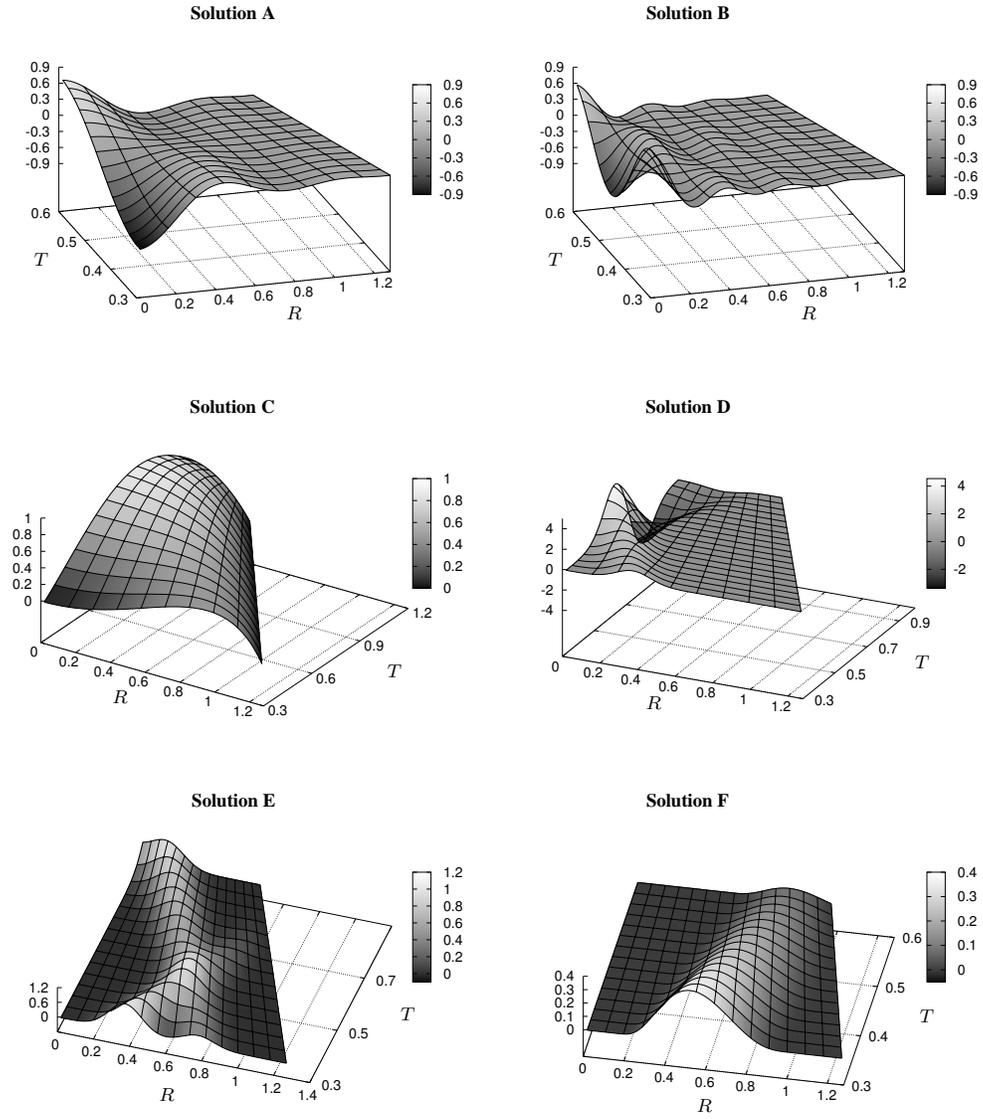,width=12.8cm}} 
  \vspace*{8pt}
  \caption{Numerical examples for the hyperboloidal IBVP of
  the linear wave equation in  Tab.~\ref{tab}, cf. Fig.~\ref{SolAF}. 
  \label{PlotsA-F}} 
\end{figure}

\newpage\mbox{}


\end{document}